\documentclass[a4paper,preprint,showpacs,preprintnumbers,amsmath,amssymb,pre]{revtex4}

\usepackage{float,epsfig,newcent,textcomp}

\begin{document}

\title{Manuscript Title:\\Recurrence Plot Based Measures of Complexity and its Application to Heart
Rate Variability Data}
\author{Norbert Marwan$^1$}
\email{email: marwan@agnld.uni-potsdam.de}
\author{Niels Wessel$^1$}

\author{Udo Meyerfeldt$^2$}
\author{Alexander Schirdewan$^2$}

\author{J\"urgen Kurths$^1$}

\affiliation{$^1$ Nonlinear Dynamics Group, Institute of Physics, University of Potsdam,
Potsdam 14415, Germany}

\affiliation{$^2$ Franz-Volhard-Hospital, HELIOS Kliniken Berlin,
Charit\'e, Humboldt University Berlin,
Wiltbergstrasse 50, 13125 Berlin, Germany}

\date{\today}

\begin{abstract}
The knowledge of transitions between regular,
laminar or chaotic behavior is essential to understand the 
underlying mechanisms behind complex systems. While several linear
approaches are often insufficient to describe such processes, there
are several nonlinear methods which however require rather long time
observations. To overcome these difficulties, we propose measures
of complexity based on vertical structures in recurrence plots and
apply them to the logistic map as well as to heart rate
variability data. For the logistic map these measures enable us 
not only to detect transitions between chaotic and periodic states, 
but also to identify laminar states, i.\,e.~chaos-chaos
transitions. The traditional recurrence quantification analysis fails
to detect the latter transitions. Applying our new measures to the
heart rate variability data, we are able to detect and
quantify the laminar phases before a life-threatening cardiac
arrhythmia occurs thereby facilitating a prediction of such an event. Our
findings could be of importance for the therapy of malignant
cardiac arrhythmias.
\end{abstract}


\maketitle

\section{Introduction}

Numerous scientific disciplines, such as astrophysics, biology or
geosciences, use data analysis techniques to get an insight 
into the complex processes observed in nature 
\cite{glass01,blasius99,marvel01} which show generally a
nonstationary and complex behavior. As these complex systems
are characterized by different transitions between regular, 
laminar and chaotic behaviors, the knowledge of these
transitions is necessary for understanding the process. However,
observational data of these systems are typically rather short. Linear
approaches of time series analysis are often not sufficient
\cite{Goldberger88,Glass93} and most of the nonlinear techniques
(cf.~\cite{abarbanel93,kantz97}), such as fractal dimensions or
Lyapunov exponents \cite{kantz97, kurths87, mandelbrot82, wolf85},
suffer from the curse of dimensionality and require rather long
data series. The uncritical application of these methods,
especially to natural data, can therefore be very dangerous and
it often leads to serious pitfalls.

To overcome these difficulties other measures of complexity have
been proposed, such as the Renyi entropies, the effective measure
complexity, the $\varepsilon$-complexity or the renormalized entropy
\cite{wackerbauer94, rapp2001}. They are mostly based on symbolic dynamics
and are efficient quantities for characterizing measurements of
natural systems, such as in cardiology \cite{kurths95,Voss96,
wessel00a}, cognitive psychology \cite{engbert97} or astrophysics
\cite{hempelmann90,schwarz93, witt94}. In this paper we focus on another type of
measures of complexity, which is based on the method of recurrence plots
(RP). This approach has been introduced for the analysis of
nonstationary and rather short data series \cite{casdagli97,
eckmann87, koebbe92}. Moreover, a quantitative analysis of
recurrence plots has been proposed to detect typical transitions
(e.\,g.~bifurcation points) occurring in complex systems \cite{trulla96,
webber94, zbilut92}. However, the quantities introduced so far are
not able to detect more complex transitions, especially
chaos-chaos transitions, which are also typical in nonlinear
dynamical systems. Therefore in this paper we introduce
measures of complexity based on recurrence plots which allow us to
identify laminar states and their transitions to regular as  well
as other chaotic regimes in complex systems. These measures make 
the investigation of intermittency of processes possible, even if they are
only represented by short and nonstationary data series.

The paper is organized as follows: After a short review of 
the technique of recurrence plots and some measures 
we introduce new measures of complexity based on
recurrence plots. After that we apply the new approach to the logistic equation
and demonstrate the ability to detect chaos-chaos transitions. Finally,
we apply this technique to heart rate variability data \cite{wessel00}.
We demonstrate that by applying our
new proposed methods we are able to detect laminar phases before the onset of a
life-threatening cardiac arrhythmia.

\section{Recurrence Plots and their Quantification}

The method of recurrence plots (RP) was firstly introduced to
visualize the time dependent behavior of the dynamics of systems, 
which can be pictured as a trajectory $\vec{x}_{i}\in 
\mathcal{R}^n$ ($i=1, \dots, N$)
in the $n$-dimensional phase space \cite{eckmann87}. It represents
the recurrence of the phase space trajectory to a certain state, which is
a fundamental property of deterministic dynamical systems
\cite{argyris94, ott93}. The main step of this visualization is
the calculation of the $N\times N$-matrix

\begin{equation}
\mathbf{R}_{i,\,j} := \Theta(\varepsilon_i-\|\vec x_{i} - \vec x_{j}\|),
\quad \, i, j=1\dots N,
\end{equation}

where $\varepsilon _{i}$ is a cut-off distance, $\Vert \cdot
\Vert$ a norm (e.\,g.~the Euclidean norm) and $\Theta (x)$
the Heaviside function.  The phase space vectors for 
one-dimensional time series $u_i$ from observations can be 
reconstructed by using the Taken's time delay method
$\vec x_i=( u_i, u_{i+\tau}, \dots, u_{i+(m-1)\,\tau})$ \citep{kantz97}.
The dimension $m$ can be estimated with the method of false
nearest neighbours (theoretically, $m=2n+1$) \citep{argyris94, kantz97}.
The cut-off distance $\varepsilon _{i}$
defines a sphere centered at $\vec{x}_{i}$. If $\vec{x}_{j}$ falls
within this sphere, the state will be close to $\vec{x}_{i}$ and
thus $\mathbf{R}_{i,j}=1$. These $\varepsilon _{i}$ can be either
constant for all $\vec{x}_{i}$ \cite{koebbe92} or they can vary in such a
way, that the sphere contains a predefined number of close states
\cite{eckmann87}. In this paper a fixed $\varepsilon_{i}$ and the
Euclidean norm are used, resulting in a symmetric RP. The
binary values in ${R}_{i,\,j}$ can be simply visualized by a
matrix plot with the colors black ($1$) and white ($0$).

The recurrence plot exhibits characteristic large-scale and
small-scale patterns which are caused by typical dynamical
behavior \cite{eckmann87,webber94}, e.\,g.~diagonals (similar
local evolution of different parts of the trajectory) or
horizontal and vertical black lines (state does not change for
some time).

Zbilut and Webber have recently developed the recurrence quantification
analysis (RQA) to quantify an RP \cite{trulla96, webber94, zbilut92}. They
define measures using the recurrence point density and the diagonal structures
in the recurrence plot, the \textit{recurrence rate}, the
\textit{determinism}, the \textit{maximal length of diagonal structures}, the
\textit{entropy} and the \textit{trend}. A computation of these
measures in small windows moving along the main diagonal of the RP
yields the time dependent behavior of these variables and, thus,
makes the identification of transitions in the time series possible \cite{trulla96}.

The RQA measures are mostly based on the distribution of the length of
the diagonal structures in the RP. Additional information about
further geometrical structures such as vertical and horizontal
elements are not included. Gao has therefore recently
introduced a recurrence time statistics, which corresponds to
vertical structures in an RP \cite{gao99, gao00}. In the following,
we extend this view on the vertical structures and define
measures of complexity based on the distribution of the vertical
line length. Since we are using symmetric RPs here, 
we will only consider the vertical structures.

\section{Measures of Complexity}

We consider a point $\vec{x}_{i}$ of the trajectory and
the set of its associated recurrence points
$S_{i}:=\{\vec{x}_{k}: \mathbf{R}_{i,\,k}
\overset{!}= 1\,; \; k \in [1 \ldots N-1]\}$. Denote a subset of
these recurrence points $s_{i}:=\{\vec{x}_{l} \in
S_{i}: (\mathbf{R}_{i,\,l} \cdot \mathbf{R}_{i,\,l+1}) + 
(\mathbf{R}_{i,\,l} \cdot \mathbf{R}_{i,\,l-1}) > 0\,;\;
l\in [1 \ldots N], \; \mathbf{R}_{i,\,0}=\mathbf{R}_{i,\,N+1}:=0\}$ 
which contains the
recurrence points forming the vertical structures in the RP at
column $i$. In continuous time systems with high time resolution
and with a not too small threshold $\varepsilon $, a large part of
this set $s_{i}$ usually corresponds to the sojourn points described
in \cite{gao99, gao00}. Although sojourn points do not occur in
maps, the subset $s_{i}$ is not necessarily empty. Next,
we determine the length $v$ of all connected subsets 
$\{ \vec{x}_{j} \not\in s_{i} ;\; \vec{x}_{j+1} \ldots 
\vec{x}_{j+v}\in s_{i} ;\; \vec{x}_{j+v+1} \not\in 
s_{i}\}$ in $s_{i}$. $P_{i}(v)=\{v_{l}\,;\; l=1,2,\dots L\}$
denotes the set of all occurring subset lengths in $s_{i}$ and from
$\bigcup_{i=1}^{N}P_{i}(v)$ we determine the distribution of the vertical line 
lengths $P(v)$ in the entire RP.

Analogous to the definition of the determinism
\cite{webber94, marwan99}, we compute the
ratio between the recurrence points forming the vertical structures and the entire
set of recurrence points
\begin{equation}
\Lambda:=\frac{\sum_{v=v_{min}}^{N}vP(v)}{\sum_{v=1}^{N}vP(v)},
\end{equation}
and call it \textit{laminarity} $\Lambda$. The computation of $\Lambda$ is
realized for $v$ which exceeds a minimal length $v_{min}$. For
maps we use $v_{min}=2$. $\Lambda$ is the measure of the
amount of vertical structures in the whole RP and represents the
occurrence of laminar states in the system, without, however, describing
the length of these laminar phases. It will decrease if the RP
consists of more single recurrence points than vertical
structures.

Next, we define the average length of vertical structures
\begin{equation}
T := \frac{\sum_{v=v_{min}}^{N} v P(v)} {\sum_{v=v_{min}}^{N} P(v)},
\end{equation}
what we call \textit{trapping time} $T$. The computation also uses
the minimal length $v_{min}$ as in $\Lambda$. The measure
$T$ contains information about the amount and the
length of the vertical structures in the RP.

Finally, we use the maximal length of the vertical structures in
the RP
\begin{equation}
V_{max}=\max\left(\{v_l\,;\; l=1,2,\dots L\}\right)
\end{equation}
as a measure, which is the analogue to the standard RQA measure
$L_{max}$ \cite{webber94}.

Although the distribution of the diagonal line
lengths also contains information about the vertical line lengths,
the two distributions are significantly different. In order 
to compare the measures proposed
with the standard RQA measures, we apply them to the logistic map.

\section{Application to the Logistic Map}

In order to investigate the potentials of $\Lambda$, $T$ and $V_{max}$, we
firstly analyze the logistic map
\begin{equation}\label{log_eq}
x_{n+1}=a\,x_{n}\left( 1-x_{n}\right),
\end{equation}
especially the interesting range of the control parameter $a\in
[3.5,4]$ with a step width of $\Delta a=0.0005$. Starting with the
idea of Trulla et al.~\cite{trulla96} to look for vertical
structures, we are especially interested in finding the 
laminar states in chaos-chaos transitions. Therefore we generate for each
control parameter $a$ a separate time series.  In the
analyzed range of $a\in [3.5,4]$ various regimes 
and transitions between them occur,
e.\,g.~accumulation points, periodic and chaotic states, band
merging points, period doublings, inner and outer crisis
\cite{collet80,oblow88,argyris94}.

\begin{figure}[htbp]
\centering \epsfig{file=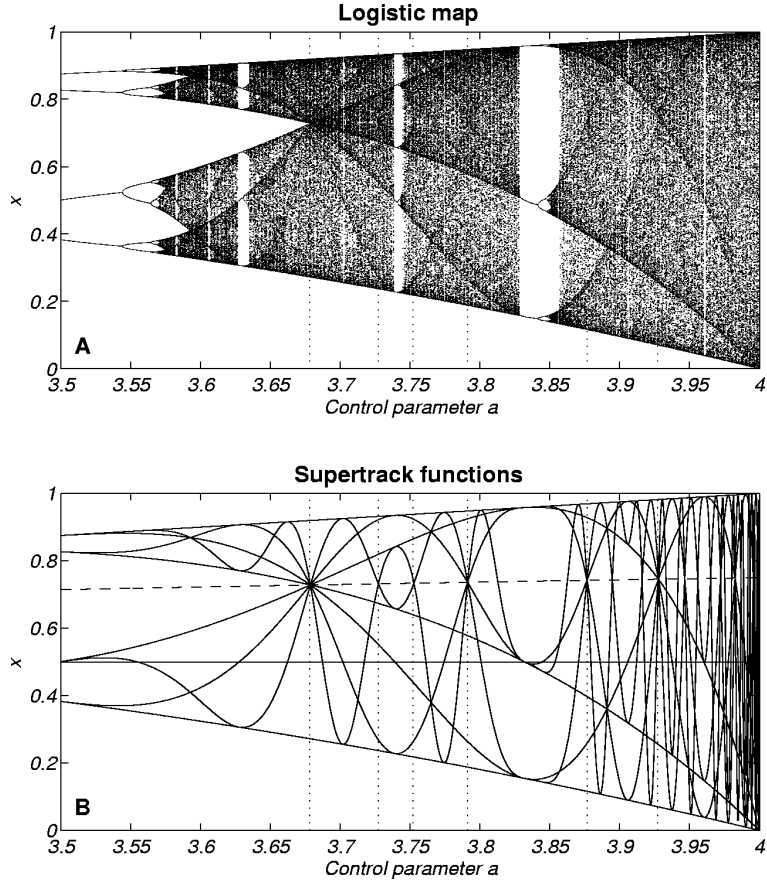, width=10cm}
\caption{(A) Bifurcation diagram of the logistic map.
(B) Low ordered supertrack functions $s_i(a)$ ($i=1\dots10$) and
the fixed point of the logistic map $1-1/a$ (dashed). 
Their intersections represent periodic windows, band merging and
laminar states. The vertical dotted lines show a choosing
of points of band merging and laminar behaviour 
($a=3.678$, $3.727$, $3.752$, $3.791$, $3.877$, $3.927$).}\label{log}
\end{figure}

A useful tool for studying the chaotic behavior are the
recursively formed \textit{supertrack functions} 
\begin{equation}\label{supertrack_eq}
s_{i+1}(a)=a \, s_{i}(a)\bigl(1-s_i(a)\bigr), \quad s_0(a)=\frac{1}{2}\,, 
\end{equation}
which
represent the functional dependence of stable states \cite{oblow88}.
The intersection of $s_{i}(a)$ with $s_{i+j}(a)$ indicates the
occurrence of a $j$-period cycle and the intersection of $s_{i}(a)$ 
with the fixed-point $(1-1/a)$ of Eq.~\ref{log_eq} indicates 
the point of an unstable singularity, i.\,e.~laminar behavior
(Fig.~\ref{log}, intersection points are marked with dotted lines). 
For each $a$ we compute a
time series of the length $N=2,000$. In order to exclude transient responses
we use the last $1,000$ values of these data series in the following
analysis.

We compute the RP after embedding the time series with a dimension
of $m=1$, a delay of $\tau =1$ and a cut-off distance of
$\varepsilon =0.1$ (in units of the standard deviation $\sigma $).
Since the considered example is a one-dimensional map, $m=1$ is 
sufficient. In general, a too small embedding leads to false 
recurrences, which are
expressed in numerous vertical structures and diagonals from the
upper left to the lower right corner \cite{gao00}. In contrast, 
an over-embedding should theoretically not distort the reconstructed phase
trajectory. Whereas false recurrences and over-embedding do
not strongly influence the measures based on diagonal structures
\cite{gao00}, the measures based on vertical structures are,
in general, much more sensitive to the embedding. This is due to the 
fact, that the embedding method causes higher order correlations 
in the phase-space trajectory, which will be of course visible in the RP.
A theoretical and more detailed explanation of this effect within 
the analysis of RPs is in preparation and beyond the scope of this 
article. For the logistic map, however,
an increasing of $m$ slightly amplifies the peaks of the vertical
based complexity measures (up to $m=3$), but it does not change 
the result significantly. The cut-off distance $\varepsilon$ is selected
to be 10 percent of the diameter of the reconstructed
phase space. Smaller values would lead to a better distinction of
small variations (e.\,g.~the range before the accumulation point
consists of small variations). However, the recurrence point
density decreases in the same way and thus the statistics of
continuous structures in the RP becomes soon insufficient. Larger
values cause a higher recurrence point density, but a lower
sensitivity to small variations.

\subsection{Recurrence Plots of the Logistic Map}

\begin{figure}[htbp]
\epsfig{file=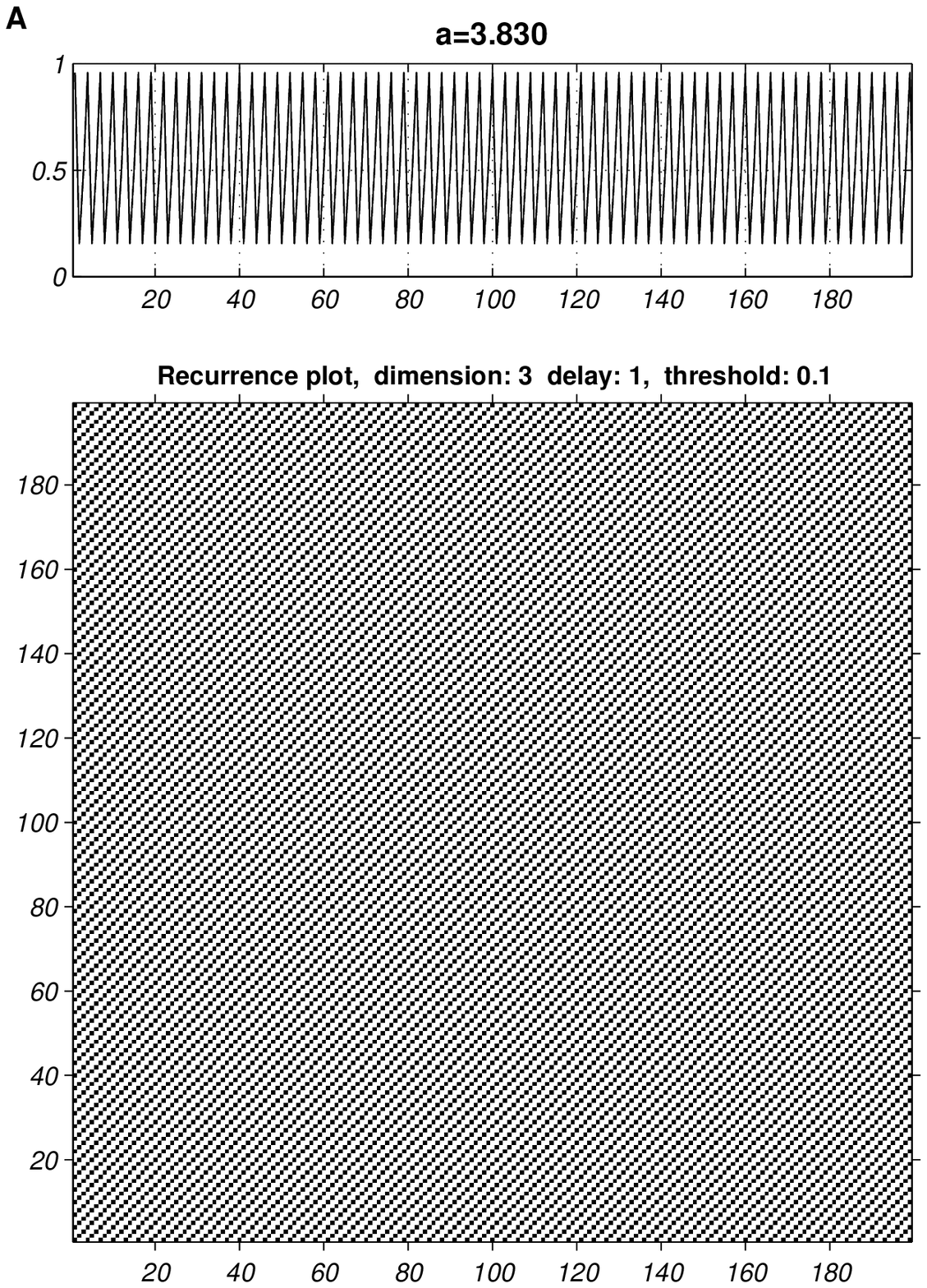, width=6cm}
\hspace*{1cm}
\epsfig{file=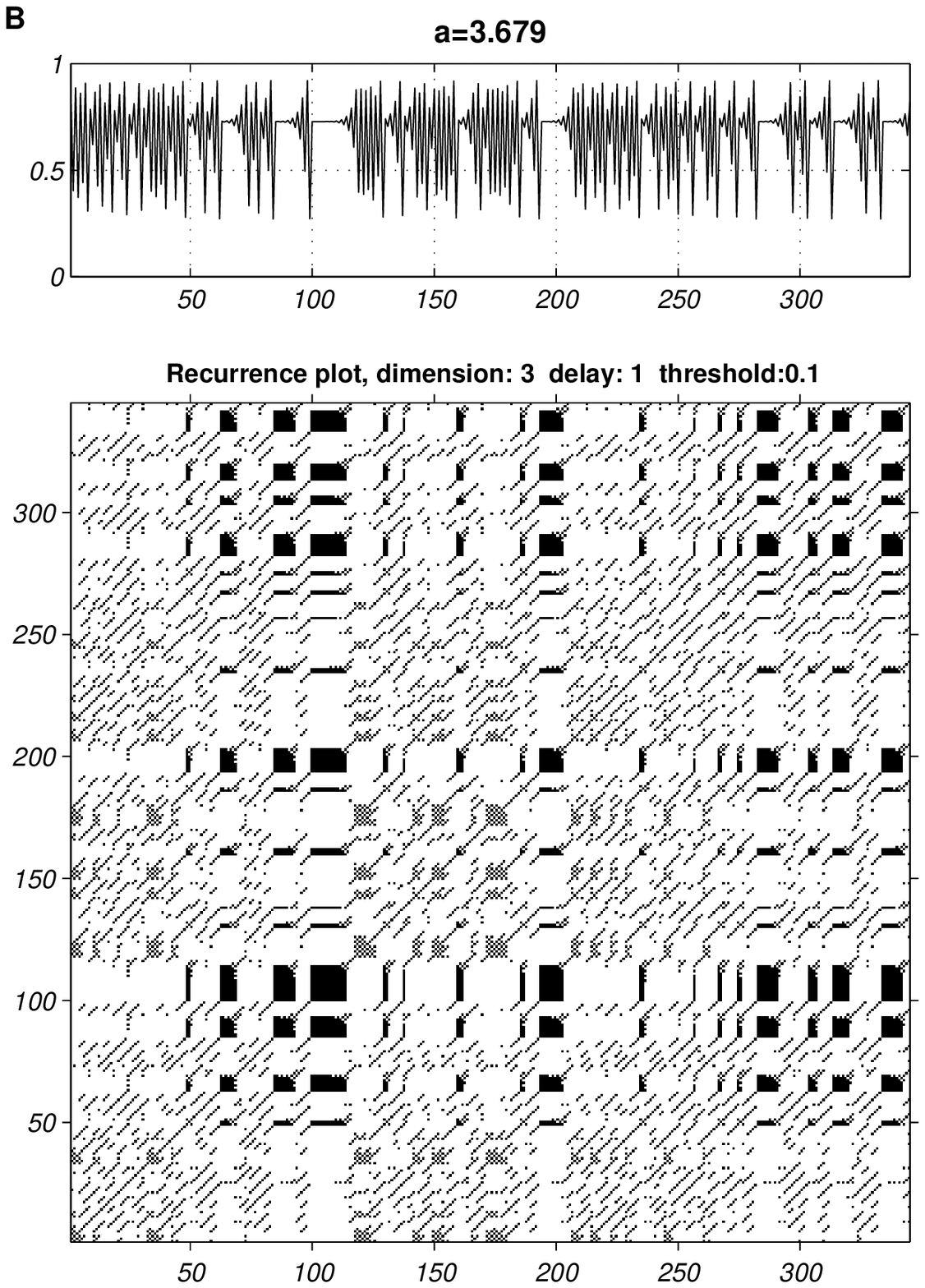, width=6cm}

\vspace{22pt}
\epsfig{file=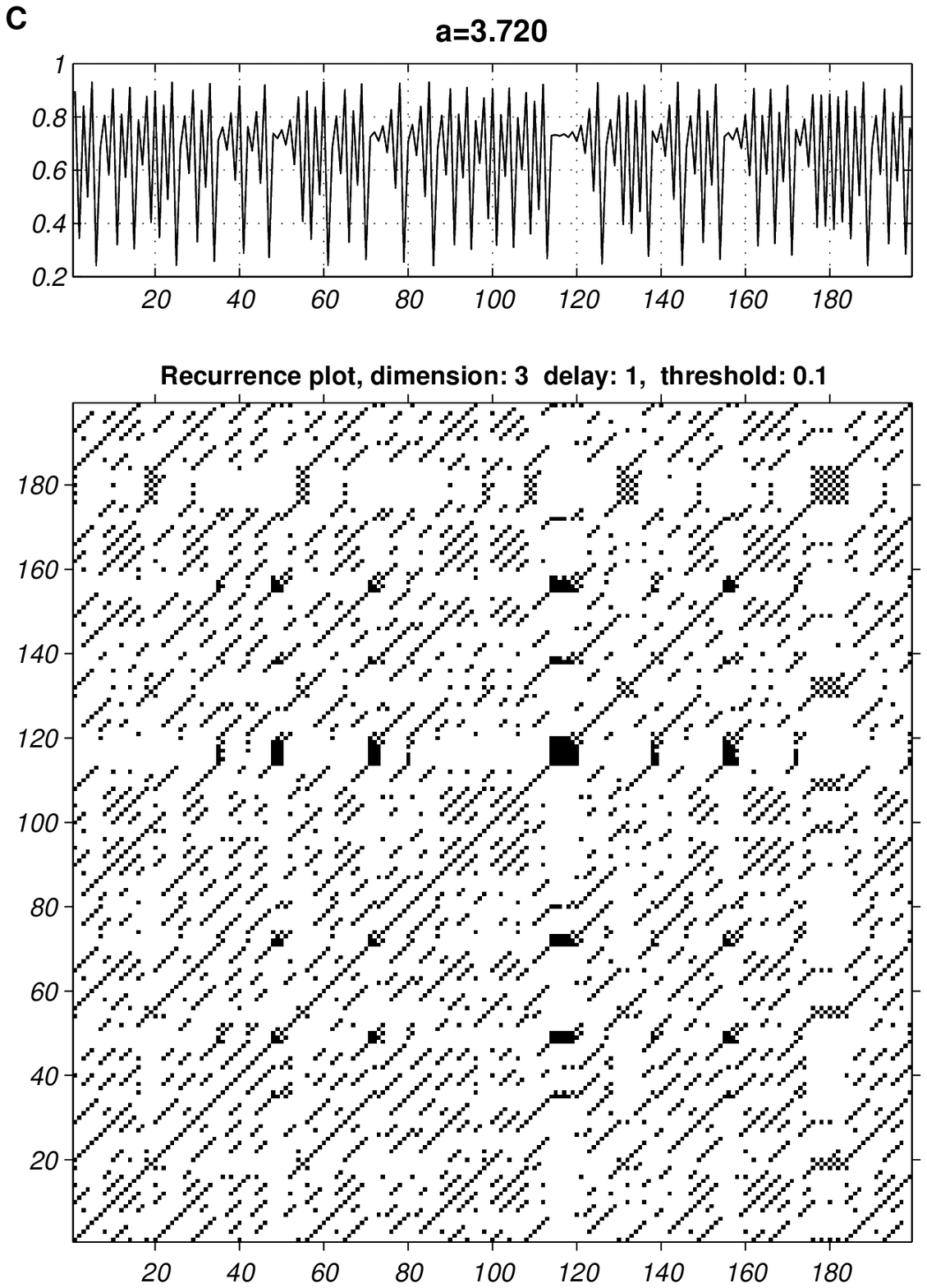, width=6cm}
\hspace*{1cm}
\epsfig{file=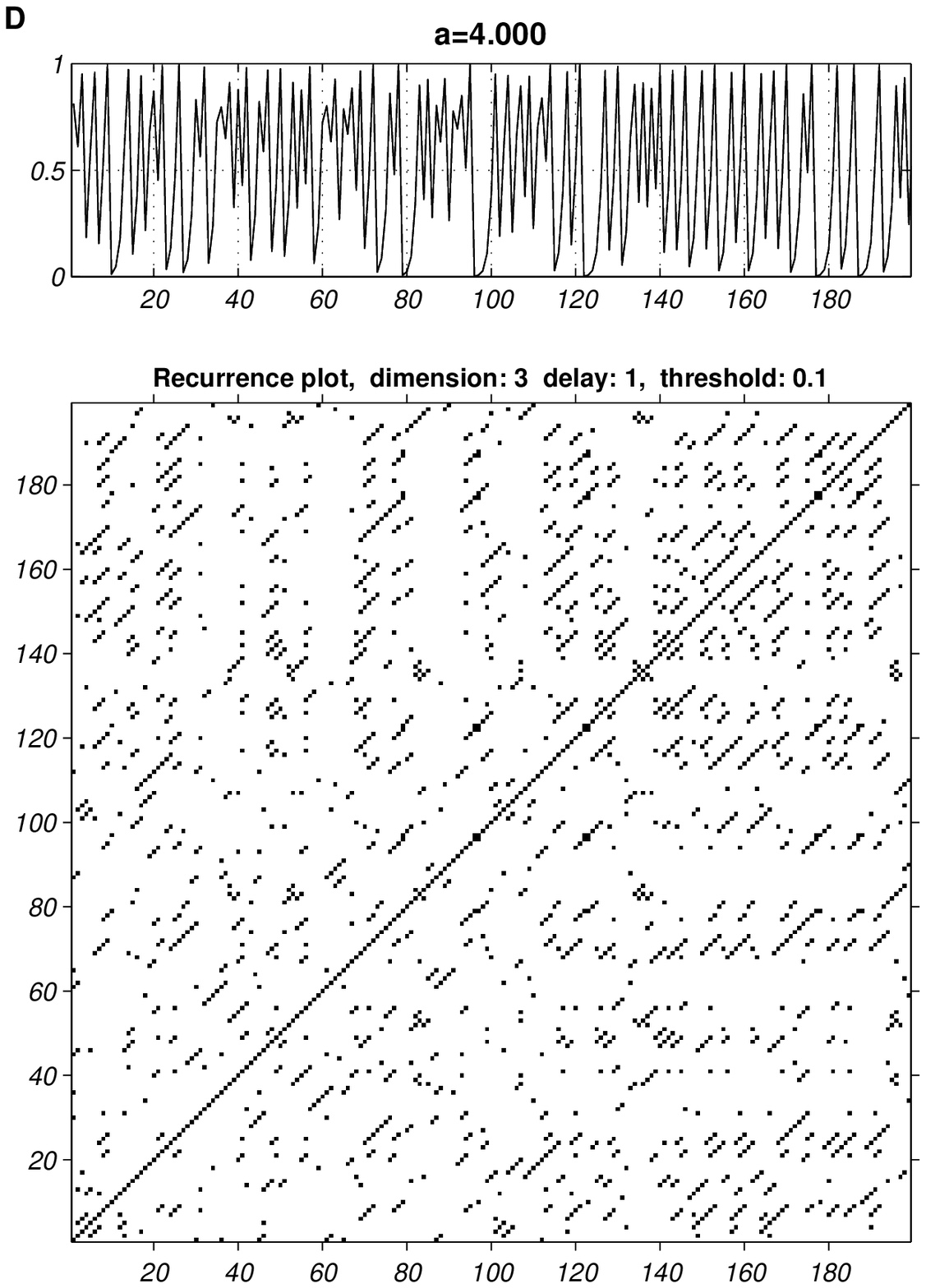, width=6cm}

\caption{Recurrence Plots (RP) of the logistic map for
various control parameters $a$, near different qualitative
changes: periodic-3-window $a=3.830$ (A), band merging $a=3.679$ (B),
supertrack intersection $a=3.720$ (C) and chaos (exterior crisis) $a=4$ (D);
with embedding dimension $m=1$, 
time delay $\tau=1$ and distance cutoff $\varepsilon=0.1\sigma$.}\label{RP}
\end{figure}

For various values of the control parameter $a$ we
obtain RPs, which already exhibit specific features
(Fig.~\ref{RP}). Periodic states (e.\,g.~in the periodic window of
length three at $a=3.830$) cause continuous and periodic
diagonal lines in the RP of a width of one. There are no vertical
or horizontal lines (Fig.~\ref{RP}\,A). Band merging points and
other cross points of supertrack functions (e.\,g.~$a=3.720$, Fig.~\ref{RP}\,C)
represent states with short laminar behavior and cause vertically
and horizontally spread black areas in the RP. The band merging
at $a=3.679$ causes frequent laminar states and therefore a lot of
vertically and horizontally spread black areas in the RP
(Fig.~\ref{RP}\,B). Fully developed chaotic states ($a=4$) cause a
rather homogeneous RP with numerous single points and rare short diagonal
or vertical lines (Fig.~\ref{RP}\,D).

\subsection{Complexity Measures of the Logistic Map}

\begin{figure}[htbp]
\centering \epsfig{file=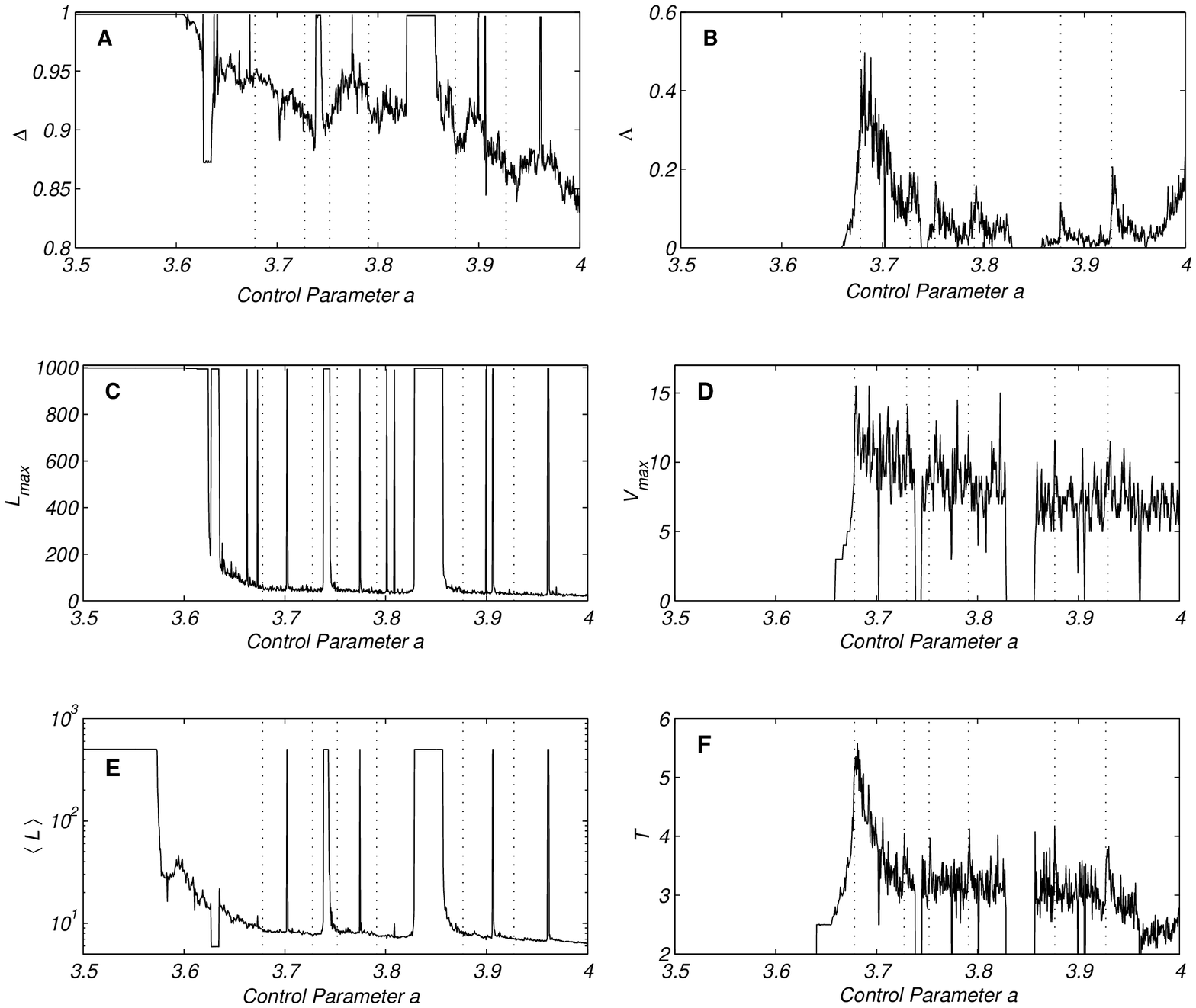, width=12.5cm}
\caption{Selected RQA parameters $\Delta$, $L_{max}$ and $\langle L \rangle$
and the new measures $\Lambda$, $V_{max}$ and $T$. 
The vertical dotted lines show some of the points of band merging 
and laminar behavior (cf.~Fig.~\ref{log}), whereby not all of them have been
marked. Whereas $\Delta$ (A), $L_{max}$ (C) and $\langle L \rangle$ (E) show 
periodic-chaotic/ chaotic-periodic transitions (maxima),
$\Lambda$ (B), $V_{max}$ (D) and $T$ (F) exhibit in addition to those 
transitions (minima) chaotic-chaotic transitions (maxima). The differences
between $\Lambda$ and $V_{max}$ are caused by the fact, that $\Lambda$ measures
only the amount of laminar states, whereas $V_{max}$ measures the
maximal duration of the laminar states. Although some peaks of $V_{max}$
and $T$ are not at the dotted lines, they correspond with laminar states
(not all can be marked).}\label{RQA}
\end{figure}

Now we compute the known RQA measures $\Delta$, $L_{max}$ and
in addition $\langle L \rangle$ (average length of diagonal lines)
and our measures $\Lambda$, $V_{max}$ and $T$ for the entire RP of
each control parameter $a$. As expected, the known RQA measures
$\Delta$, $L_{max}$ and $\langle L \rangle$ clearly detect the
transitions from chaotic to
periodic sequences and vice versa (Fig.~\ref{RQA}\,A, C, E)
\cite{trulla96}. However, it seems that one cannot get more
information than the periodic-chaotic/ chaotic-periodic transitions.
Near the supertrack crossing points (band merging points
included), e.\,g.~$a=3.678, 3.791, 3.927$, there
are no significant indications in these RQA measures. They
clearly identify the bifurcation points (periodic-chaotic/
chaotic-periodic transitions), without, however, finding the
chaos-chaos transitions and the laminar states.

Calculating the vertical based measures $\Lambda$ and $T$,
we are able to find the periodic-chaotic/
chaotic-periodic transitions and the laminar states 
(Fig.~\ref{RQA}\,B, F). The
occurrence of vertical lines starts shortly before the band
merging from two to one band at $a=3.678\dots $

For smaller $a$-values the consecutive points jump between the two
bands and it is therefore impossible to obtain a laminar behavior. 
A longer persistence of
states is not possible until all bands are merged. However, due to
the finite range of neighborhood searching in the phase space,
vertical lines occur before this point. 

Vertical lines occur much more frequently at supertrack crossing
points (band merging points included), than in other chaotic
regimes, which is revealed by $\Lambda$ (cf.~Fig.~\ref{RQA}\,B, again,
supertrack crossing points are marked with dotted lines). As in
the states before the merging from two to one band, vertical
lines are not found within periodic windows, e.\,g.~$a=3.848$. The
mean of the distribution of $v$ is the introduced measure $T$
(Fig.~\ref{RQA}\,F). It will vanish if $a$ is smaller than the
point of merging from two to one band. $T$ increases at those
points where more low ordered supertrack functions are crossing
(Fig.~\ref{RQA}\,F). This corresponds to the occurrence of
laminar states. Although $V_{max}$ also reveals laminar states, it
is quite different from the other two measures, because it gives
the maximum of all of the durations of the laminar states. However,
periodic states are also associated with vanishing $T$ and
$V_{max}$.

Hence, the vertical length based measures yield periodic-chaotic/
chaotic-periodic as well as chaos-chaos transitions (laminar
states).

We have also computed $\Lambda$, $V_{max}$ and $T$ for the logistic
map with transients using the same approach as described in
\cite{trulla96}. The qualitative statement of the measures is the
same as above.

\section{Application to heart rate variability data}

Heart rate variability (HRV) typically shows a complex 
behavior and it is difficult to identify disease specific patterns
\cite{schumann2002}.
A fundamental challenge in cardiology is to find early signs of
ventricular tachy{\-}arrhythmias (VT) in patients with an
implanted cardioverter-defibrillator (ICD) based on HRV data 
\cite{diaz01, Huikuri01, wessel00, Guzzetti01}. Therefore 
standard HRV parameters from time and
frequency domain \cite{TaskForce1996}, parameters from symbolic
dynamics \cite {kurths95, Voss96} as well as the finite-time
growth rates \cite{nese89} were applied to the data of a clinical
pilot study \cite{wessel00}. Using two nonlinear approaches, we have
recently found significant differences between control and VT time
series based mainly on laminar phases in the data before a VT.
Therefore the aim of this investigation is to test whether our RP
approach is suitable to identify and quantify these laminar
phases.

The defibrillators used in the study cited (PCD 7220/7221, Medtronic)
are able to store at least 1000 beat-to-beat intervals prior to
the onset of VT (10 ms resolution), corresponding to approximately
9--15 minutes. We reanalyze these intervals from 17 chronic heart
failure ICD patients just before the onset of a VT and at a
control time, i.\,e.~without a following arrhythmic event. Time
series including more than one non-sustained VT, with induced VT's,
pacemaker activity or more than 10\,\% of ventricular premature
beats are not considered in this analysis. Some patients had
several VT's; we finally had 24 time series with a subsequent VT
and the respective 24 control series without a life-threatening
arrhythmia. In order to analyze only the dynamics occurring just before a
VT, the beat-to-beat intervals of the VT itself at the end of the
time series are removed from the tachograms.

\begin{figure}[htbp]
\centering 
\mbox{\epsfig{file=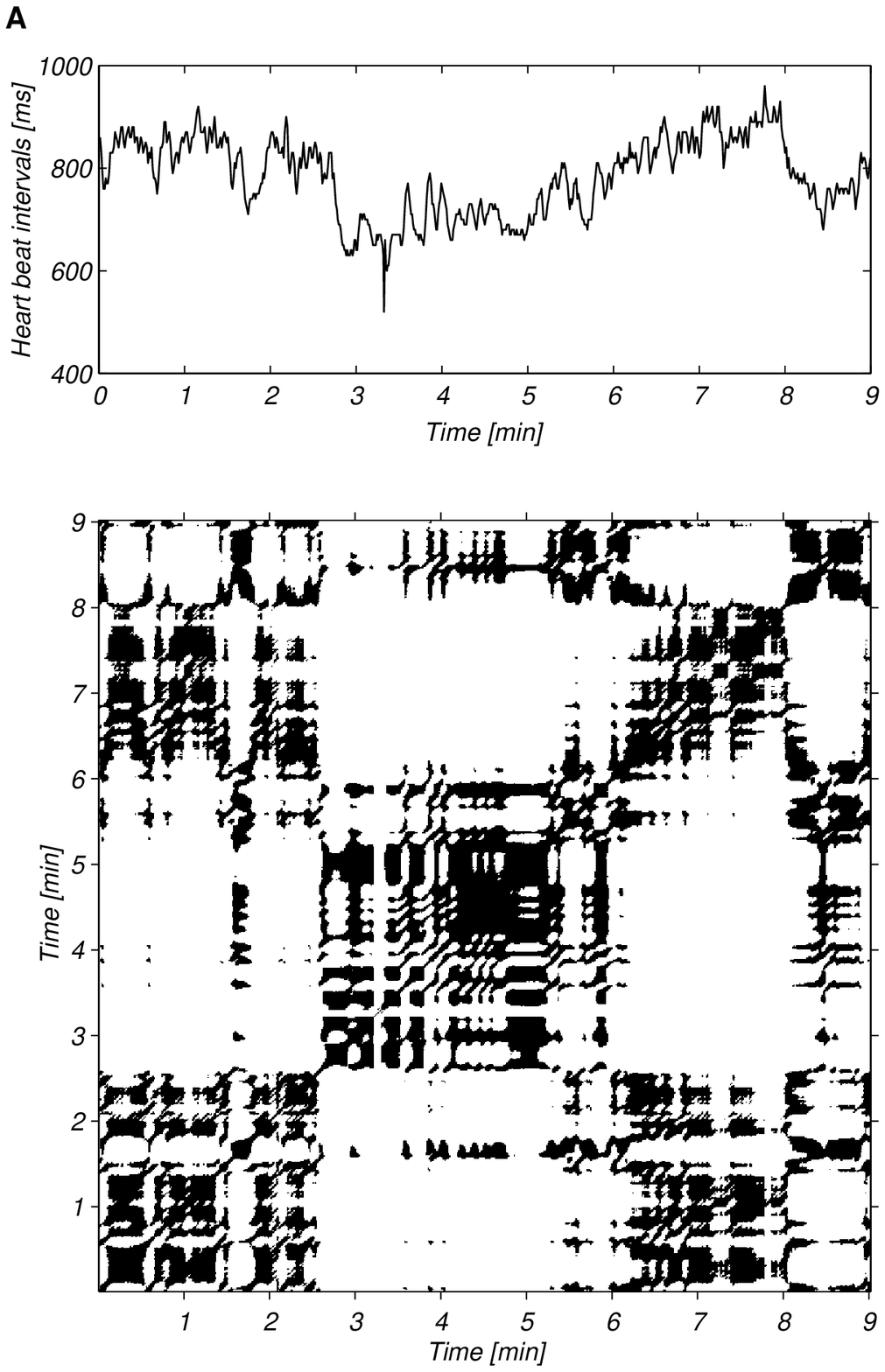, width=6cm}
\hspace*{1cm}
\epsfig{file=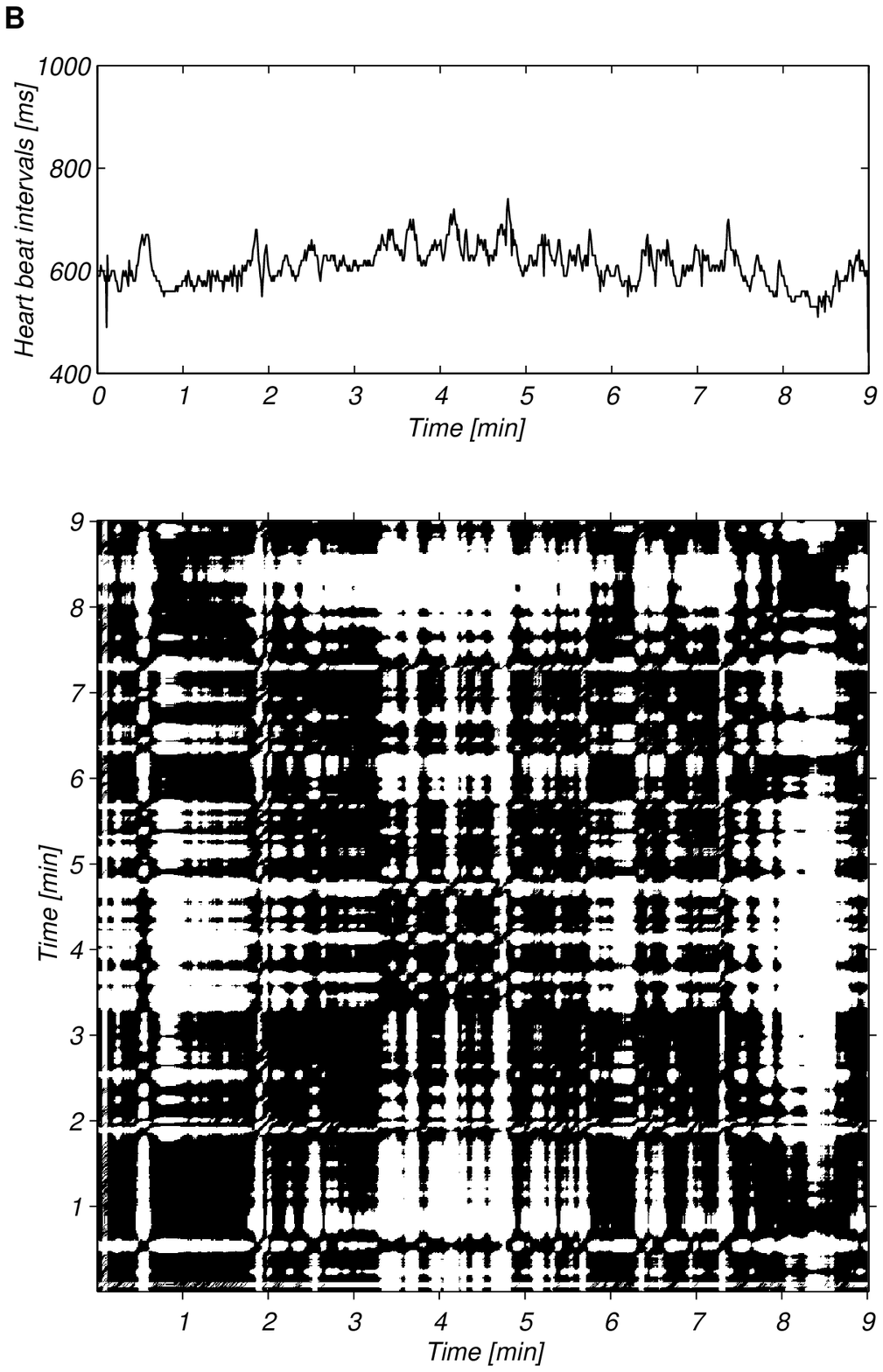, width=6cm}}
\caption{Recurrence plots of the heart beat interval time series
at a control time (A) and before a VT (B)
with an embedding of 6 and a radius of 110. The RP before a
life-threatening arrhythmia is characterized by big black
rectangles whereas the RP from the control
series shows only small rectangles.}\label{RP_herz}
\end{figure}

We calculate all standard RQA parameters described in
\cite{webber94} as well as the new measures laminarity $\Lambda$,
trapping time $T$ and maximal vertical line length $V_{max}$ (in similarity
to the maximal diagonal line length $L_{max}$) for
different embedding dimensions $m$ and nearest neighbouring radii
$\varepsilon$. We find differences between both groups of data for several of
the parameters mentioned above. However, the most significant parameters
are $V_{max}$ and $L_{max}$ for rather large radii (Tab.~\ref{tablerqa}).
The vertical line length $V_{max}$ is more powerful in
discriminating both groups than the diagonal line length
$L_{max}$, as can be recognized by the higher $p$-values for $V_{max}$
(Tab.~\ref{tablerqa}). Figure~\ref{RP_herz} gives a
typical example of the recurrence plots before a VT and at a
control time with an embedding of 6 and a radius of 110. The RP
before a life-threatening arrhythmia  is characterized by large
black rectangles ($V_{max}=242$ here), whereas the RP from the
control series shows only small rectangles ($V_{max}=117$).

\begin{table}[t]
\centering
\caption{Results of maximal diagonal and vertical line length shortly
before VT and at control time, nonparametric Mann-Whitney U-test,
$p$ -- significance; * -- $p<0.05$; ** -- $p<0.01$;
n.\,s. -- not significant $ p\geq 0.05$) \label{tablerqa}}
\begin{tabular}{r@{\hspace{8pt}}r@{\hspace{16pt}}r@{\hspace{16pt}}r@{\hspace{12pt}}l}
\hline
$m$&$\varepsilon$& \multicolumn{1}{c}{VT\hspace*{12pt}} & \multicolumn{1}{c}{Control\hspace*{12pt}} & $p$ \\
\hline
\hline
\multicolumn{4}{l}{{\scriptsize {\it Maximal diagonal line length $L_{max}$}}} \\
$3$&$77$ & {396.6$\pm $253.8} & {261.5$\pm $156.6} & {n.\,s.} \\
$6$&$110$ & {447.6$\pm $269.1} & {285.5$\pm $160.4} & {*} \\
$9$&$150$ & {504.6$\pm $265.9} & {311.6$\pm $157.2} & {*} \\
$12$&$170$ & {520.7$\pm $268.8} & {324.7$\pm $180.2} & {*} \\
\hline
\multicolumn{4}{l}{{\scriptsize {\it Maximal vertical line length $V_{max}$}}}\\
$3$&$77$ & {261.4$\pm $193.5} & {169.2$\pm $135.9} & {*} \\
$6$&$110$ & {283.7$\pm $190.4} & {179.5$\pm $134.1} & {**} \\
$9$&$150$ & {342.4$\pm $193.6} & {216.1$\pm $137.1} & {**} \\
$12$&$170$ & {353.5$\pm $221.4} & {215.1$\pm $138.6} & {**} \\
\hline
\end{tabular}
\end{table}

\section{Summary}

We have introduced three new recurrence plot (RP) based measures of complexity,
the laminarity $\Lambda$, the trapping time $T$ and the maximal length
of vertical structures in the RP $V_{max}$.
These measures of complexity have been applied to the logistic map
and heart rate variability data. In contrast to the known RQA
measures (\cite{trulla96}, \cite{zbilut92}),
which are able to detect transitions between chaotic and periodic
states (and vice versa), our new measures enable us to identify
laminar states too, i.\,e.~chaos-chaos transitions.
These measures are provided
by the vertical lines in recurrence plots. The occurrence of vertical (and
horizontal) structures is directly related to the occurrence of laminar
states.

The laminarity $\Lambda$ enables us generally to detect laminar states in
a dynamical system. The trapping time $T$ contains information
about the frequency of the laminar states and their lengths. The
maximal length $V_{max}$ reveals information about the
time duration of the laminar states thus making the 
investigation of intermittency possible.

If the embedding of the data is too small, it will lead to false recurrences,
which is expressed in numerous vertical structures and diagonals perpendicular 
to the main diagonal. Whereas false recurrences do
not influence the measures based on diagonal structures,
the measures based on vertical structures are
sensitive to it.

The application of these measures to the logistic equation for a
range of various control parameters has revealed points of laminar
states without any additional knowledge about the characteristic
parameters or dynamical behavior of the specific systems. Nevertheless,
$\Lambda$, $V_{max}$ and $T$ are different in their magnitudes.
Further investigations are necessary to understand all relations
between the magnitudes of $V_{max}$ and the recognized
chaos-chaos transitions.

The application of the new complexity measures to the ICD stored
heart rate data before the onset of a life-threatening arrhythmia
seems to be very successful for the detection of laminar phases thus
making a prediction of such VT possible. The differences between
the VT and the control series are more significant than in
\cite{wessel00}. However, two limitations of this study are
the relative small number of time series and the reduced
statistical analysis (no subdivisions concerning age, sex and
heart disease). For this reason, our results should to be
validated on a larger data base. Furthermore, this investigation
could be enhanced for tachograms including more than 10\%
ventricular premature beats. In conclusion, this study has
demonstrated that the RQA based complexity measures could play an
important role in the prediction of VT events even in short term HRV
time series.

Many biological data contain epochs of laminar states,
which can be detected and quantified by the RP based measures. We have
demonstrated differences between the measures based on 
the vertical and the diagonal structures and
therefore we suggest the use of the method proposed in this article
in addition to the traditional measures.

A download of the Matlab implementation
is available at: www.agnld.uni-potsdam.de/\texttildelow marwan.

\section{Acknowledgments}
This work was partly supported by the priority programme SPP\,1097 
of the German Science Foundation (DFG).
We gratefully acknowledge M.~Romano, M.~Thiel and U.~Schwarz 
for fruitful discussions.

\bibliography{../mybibs,herzbibs}

\printfigures

\end{document}